\documentstyle[psfig]{l-aa}

\def\spose#1{\hbox to 0pt{#1\hss}}
\def\simlt{\mathrel{\spose{\lower 3pt\hbox{$\mathchar"218$}}
     \raise 2.0pt\hbox{$\mathchar"13C$}}}
\def\simgt{\mathrel{\spose{\lower 3pt\hbox{$\mathchar"218$}}
     \raise 2.0pt\hbox{$\mathchar"13E$}}}

\begin{document}
\thesaurus{ }
\title{Light and Heavy Elements in the Galactic Bulge}
\author{Francesca Matteucci$^{1,2}$, Donatella Romano$^{1,2}$, 
Paolo Molaro$^{3}$}
\offprints{F. Matteucci}
\institute{
Dipartimento di Astronomia, Universit\`a di Trieste, Via G.B. Tiepolo, 11,
34131 Trieste, Italy \and
SISSA/ISAS, Via Beirut 2-4, 34014 Trieste, Italy \and
Osservatorio Astronomico di Trieste,
Via G.B. Tiepolo, 11, 34100 Trieste, Italy}
\maketitle

\begin{abstract}
In the context of an inside-out model for the 
formation of our Galaxy we present 
results for the chemical evolution of the Galactic bulge 
by assuming that this central region
evolved even 
faster than the Galactic halo. 
This assumption is required in order to reproduce the observed 
metallicity distribution of bulge stars as obtained by
McWilliam \& Rich (1994).\\
The model is similar to that adopted by Matteucci and Brocato (1990)
with the exception that
we have adopted the most recent nucleosynthesis prescriptions 
for either low-intermediate mass stars or massive stars
and extended our predictions to most of the $\alpha$-elements, Fe, C, N 
and light elements (D, $^{7}$Li). 
We have tested the effect of changing the slope of the IMF 
on the predicted stellar metallicity distribution 
for bulge stars and have compared our results with the 
distribution obtained from the most recent data. 
An initial mass function (IMF) favoring the formation of 
massive stars with respect to the solar vicinity
improves the agreement with observations. 
Then, using our best model, we have made some predictions 
about temporal evolution of light and heavy 
species in the Galactic bulge. 
In particular, we predict that $\alpha$-elements should be enhanced 
relative to Fe for most of the metallicity
range, although different $\alpha$-elements should 
show a different degree of enhancement due to the particular
nucleosynthetic history of each element.
The enhancements are larger than predicted for the halo and thick-disk stars
because of the flatter IMF assumed for the bulge.
We predict that deuterium, due to the intense star formation 
occurred in bulge, should be completely depleted 
whereas $^{7}$Li should show a trend similar to that found for 
the solar neighbourhood. We also compared the 
Li abundance recently measured in one bulge
star (Minniti et al.\,, 1998) and concluded that it is very likely that the Li 
in the star must have been depleted and therefore it does not
reflect the Li abundance of the gas out of which the star formed.
In the opposite case, one should completely suppress one of the 
main stellar sources of Li in order to reconcile the model
with the observed value.
\end{abstract}

\section{Introduction}
In recent years, several studies dealing with a detailed 
spectroscopic analysis of K giants in Baade's window 
have appeared in literature 
(Rich, 1983, 1986, 1988; McWilliam \& Rich, 1994) 
and have 
presented a stellar metallicity distribution of bulge K giants with a shape 
different from that of the G-dwarf distribution for the solar neighbourhood. 
In particular, McWilliam \& Rich (1994) found that the metallicity peak 
of the distribution 
of bulge stars appears at a metallicity higher than for 
the solar neighbourhood, 
at around [Fe/H] $\sim$ +\,0.0 dex. New data on the 
G-dwarf distribution for the solar 
vicinity (Rocha-Pinto \& Maciel, 1996) show, in fact, a 
well-defined peak in metallicity between [Fe/H] = -\,0.3 and +\,0.0.\\
The stellar metallicity distribution is an important constraint 
for chemical evolution models, because it is representative of 
the chemical enrichment of the specific Galactic component it is referring to.
In fact, the stars used to give such a distribution have lifetimes 
greater than or equal to the age of the Galaxy, hence providing 
a complete picture of the chemical enrichment history for the considered 
Galactic component. 
This is because different stellar distributions versus metallicity are likely 
to be due to different time scales of collapse 
and different star formation histories.\\
Matteucci \& Brocato (1990) predicted that the ratio of some 
$\alpha$-elements (Si, Mg and O) 
to Fe in the bulge stars should be
larger than solar over most of the metallicity range
and similar to what is observed in halo stars. 
This was due to the assumption of a fast evolution for the bulge 
leading the gas to be fastly enriched in Fe by essentially supernovae 
of type II.
Later McWilliam \& Rich (1994) showed that indeed Mg and Ti are 
enhanced by $\sim 0.3$ dex relative
to Fe  over the whole [Fe/H] range, in very good agreement with 
Matteucci \& Brocato (1990).
However, they also found that elements such as Ca and Si closely 
follow the trend of disk stars, namely they reach solar ratios 
at [Fe/H] $>$ -\,0.2 dex, 
at variance with the predictions.
More recently, Sadler et al.\,(1996) found Mg enhancements 
($\sim 0.2$ dex) 
only for bulge 
stars with [Fe/H] $<$ 0.0, whereas for stars with larger metallicities 
they found [Mg/Fe] = 0.0.
These discrepancies suggest that 
more detailed abundances are necessary to 
assess this point and draw firm conclusions on
the bulge formation mechanism.
In particular, we would like to assess if the bulge formed  very fast
reaching a high metallicity in a short timescale (Rich, 1988; Renzini, 1993,
1994) or if instead it formed by merging of bulgeless spirals on a 
Hubble time (Schweizer \& Seitzer, 1988). In favor of the first hypothesis
there are the old ages of stars in the bulge as derived by Terndrup (1988)
as well as the old ages of bulge globular clusters (Ortolani et al.\,, 1995),
all suggesting that the bulge formed contemporarily or perhaps 
even before the halo (but see Holtzman et al. \,,1993).
Abundance ratios represent an independent constraint for the 
mechanism of bulge formation since they can be very different
according to rapid or slow star formation.
The aim of this paper is to study 
the behaviour of a large number of abundances and abundance ratios 
in the bulge under the assumption of a fast bulge formation process.\\

In particular, we have used a model for the chemical evolution of the bulge
which allows us to 
follow the temporal evolution of twenty chemical and isotopic 
species (H, $^{2}$H, $^{3}$He, $^{4}$He, $^{7}$Li, 
$^{12}$C, $^{13}$C, $^{14}$N, $^{16}$O, neutron rich species, 
$^{20}$Ne, $^{24}$Mg, $^{28}$Si, $^{32}$S, $^{40}$Ca, 
$^{56}$Fe, Ni, Cu, Zn, Kr).
The model is an extension to the bulge of the model of Chiappini et al. (1997).
In this model the halo-thick disk and the thin disk form out of two distinct infall episodes. In particular, the halo forms during the first infall
episode on a time scale of roughly 1 Gyr whereas the
thin disk forms out of extragalactic gas on long timescales 
increasing with galactocentric distance (inside-out formation). 
Here we assume that the bulge formed out of the same gas forming the halo 
(Wyse and Gilmore, 1992) but 
it evolved faster than the halo itself being the more condensed 
central part of the Galactic spheroid. 
The bulge is represented by a region with a 2 Kpc radius 
and a total mass of about $10\,^{10}$ M$_\odot$. 
Possible exchange of matter between the bulge and the 
Galactic disk is not taken into account.\\
We have also followed in detail the evolution of the $^{7}$Li abundance 
in the bulge by assuming several $^{7}$Li stellar sources in 
agreement with the results 
of Matteucci et al.\,(1995). These predictions are also important since
very recently the advent of microlensing has given us the 
opportunity of observing the faintest objects 
in our Galaxy and in particular in the bulge. 
Minniti et al.\,(1998) have measured 
the lithium abundance in the atmosphere 
of a bulge star located near the main sequence 
turn-off called 97\,BLG\,45 and claimed that the derived Li abundance 
reflects the primordial one. This fact, if true, is extremely interesting 
since all the halo stars show the same Li abundance 
(Li-plateau, see Bonifacio \& Molaro, 1997) 
which is generally interpreted as being the primordial one.\\
In paragraph 2 we present the chemical evolution model, 
in paragraph 3 we show the main results 
we have obtained (compared, if possible, with the observations) 
and finally, in paragraph 4, we draw some conclusions.

\section{The chemical evolution model}
The basic assumptions of the bulge model are:
\begin{itemize}
\item[1)] Instantaneous mixing of gas.
\item[2)] No instantaneous recycling approximation.
\item[3)] A star formation rate (SFR) expressed as:
\begin{displaymath}
\psi(r,t) = \nu\,\left( \frac{\sigma(r,t)}{\tilde{\sigma}(\tilde{r},t)} \right) ^{2\,(k - 1)}\,\cdot
\end{displaymath}
\begin{displaymath}
\cdot\,\left( \frac{\sigma(r,t_{gal})}{\sigma(r,t)} \right) ^{k - 1}\,G^{k}(r,t)\,;
\end{displaymath}
where $\tilde{\sigma}(\tilde{r},t)$ is the total surface mass density at a radius 
$\tilde{r} = 10$ Kpc, $k = 1.5$ is suggested as the best value for the solar neighbourhood 
by Chiappini et al.\,(1997), $\nu = 20$ Gyr$\,^{-1}$ is the star formation efficiency in the bulge.
This value is similar to what is usually assumed for an elliptical galaxy of the same mass
(Matteucci and Tornamb\'e, 1987) since bulges of spirals are very similar to ellipticals for what concerns their stellar content and fall in the same region of the fundamental plane (Jablonka et al.\,, 1996).\\
For the solar vicinity $\nu$ is set equal to 2 Gyr$\,^{-1}$ in 
the halo and thick disk components and 1 Gyr$\,^{-1}$ in the thin 
disk component. This choice is made for the models to be in agreement 
with the observational constraints and follows the prescriptions of 
the Chiappini et al. (1997) model.\\
$G(r,t) = \sigma_{g}(r,t)/\sigma(r,t_{gal})$ is the normalized surface 
gas density, $t_{gal} = 15$ Gyr is the Galactic lifetime and $\sigma$
is the total surface mass density.
\item[4)] An IMF expressed as a power law, 
with index $x_{IMF}$:
\begin{displaymath}
\varphi(m) \propto m^{\,- \, (1 + x_{IMF})}\,.
\end{displaymath}
We have explored four cases: ({a}) $x_{IMF} = 1.35$ (Salpeter, 1955), ({b}) $x_{IMF} = 1.1$ (Matteucci \& Brocato, 1990), 
({c}) $x_{IMF} = 0.95$ (Matteucci \& Tornamb\'e, 1987) 
for the whole mass range (0.1 -- 100 M$_\odot$) and ({d}) 
$x_{IMF} = 1.35$ for 0.1 -- 2 M$_\odot$ and $x_{IMF} = 1.7$ 
for M $>$ 2 M$_\odot$ (Scalo, 1986).
\item[5)] A gas collapse rate expressed as:
\begin{displaymath}
\frac{dG(r,t)_{infall}}{dt} = \frac{A(r)}{\sigma(r,t_{gal})}\,e^{-t/\tau}\,,
\end{displaymath}
where $\tau = 0.1$ Gyr for the bulge, $\tau =8$ Gyr 
for the solar vicinity and $\tau = 1$ Gyr for the halo.
$A(r)$ is derived from the condition of reproducing the 
current total surface mass density distribution in the bulge.
\item[6)] We adopted detailed nucleosynthesis prescriptions
for the $\alpha$-elements and Fe from type I SNe and type II SNe as in 
Chiappini et al.\,(1997).
The nucleosynthesis of He, C and N from low and intermediate mass stars 
(0.8 --8 M$_\odot$)
is taken from van den Hoek and Groenewegen (1997) (their standard model).
We remind that $^{13}$C and $^{14}$N are partly {\it secondary} 
and partly {\it primary}
in low and intermediate mass stars whereas they are only 
secondary in massive stars.
The D is assumed to be completely destroyed and the prescriptions for
the production of $^{3}$He are the same as in Chiappini et al. (1997).
The nucleosynthesis prescriptions for Lithium are described in detail 
in paragraph 3.3.2.
Finally, the nucleosynthesis of Cu, Zn, Ni and Kr is the same as 
in Matteucci et al.\,(1993).
\item[7)] The possibility of Galactic winds is not taken into account
since they seem not to be appropriate for our Galaxy (Tosi et al.\,, 1998).
\end{itemize}

\begin{figure}
\centering
\vspace{12.0cm}
\caption[]{\label{fig:fig 1} Predicted stellar metallicity distributions 
normalized to the total number of stars N$_{tot}$ (continuous lines)
compared to the recent data sample of McWilliam \& Rich (1994) 
(dashed lines).}
\end{figure}

\begin{table}
\begin{center}
\scriptsize{\textbf{Tab.1.} Model inputs.}
\paragraph{}
\begin{tabular}{c|c c c c} 
\multicolumn{5}{c}{}\\\hline
\\
\footnotesize{Model}&\footnotesize{C-stars}&\footnotesize{M-AGB}&\footnotesize{SNeII}&\footnotesize{$x_{IMF}$}\\
\\\hline
\\
$1$&log N$_{Li} = 3.85$&log N$_{Li} = 4.15$&WW95&1.35\\
\\
$2$&log N$_{Li} = 3.85$&log N$_{Li} = 4.15$&no&1.35\\
\\
$3$&log N$_{Li} = 3.85$&log N$_{Li} = 3.5$&no&$1.35$\\
\\
$4$&log N$_{Li} = 3.85$&log N$_{Li} = 3.5$&no&1.1\\
\\\hline
\end{tabular}
\end{center}
\end{table}

\section{Model results}
\subsection{Stellar metallicity distribution}
We have studied four cases: $x_{IMF} = 1.35$, $x_{IMF} = 1.1$, 
$x_{IMF} = 0.95$ for the whole mass range as well as an IMF 
like that suggested 
by Scalo (1986). As shown in Fig.1, a decrease in the power 
law index of the IMF moves 
the peak of the predicted stellar distribution towards higher metallicity
values. The observed distribution (McWilliam \& Rich, 1994) 
is shown for comparison. The metallicity distribution of bulge K giants 
is well reproduced by an IMF with a power index in the range 1.1 -- 1.35. 
This claim is in agreement with Matteucci \& Brocato (1990), 
who suggested a value of $x_{IMF}$ in the range 1.1 -- 1.26 
(that is a lower power index than the Salpeter one).\\
At this point it is worth emphasizing that while Matteucci \& Brocato 
used as a constraint the observed metallicity distribution of bulge 
K giants of Rich (1988), here we adopt the more recent one 
of McWilliam \& Rich (1994), which differs from the previous one 
in the position of 
the metallicity peak which moves from +\,0.35 dex to +\,0.0 dex.\\
Concerning the solar neighbourhood, in their paper Chiappini et al.\,(1997) 
have stressed that the metallicity distribution of G dwarfs can be 
well reproduced by a long ($\sim 8$ Gyr) time scale 
of formation and a moderate SFR. Moreover they have adopted a Scalo IMF; 
this produces results consistent with both G dwarf distribution and 
abundance ratios constraints.

\begin{figure}
\centering
\vspace{12.0cm}
\caption[]{\label{fig:fig 2} The predicted [el/Fe] vs [Fe/H] relations for
$^{12}$C, $^{13}$C and $^{14}$N in the Galactic bulge (continuous lines) and 
in the solar neighbourhood (dashed lines).
The predictions for the solar neighbourhood are obtained by means of the 
model of Chiappini et al. (1997).
The assumed solar abundances are those of Anders \& Grevesse (1989).
The differences can be explained as a consequence of the different 
age-metallicity relations characterizing the bulge and the solar 
vicinity region, due to faster evolution of the central region. 
Moreover, a non negligible effect on the abundance ratios is given 
by the presence of a larger number of massive stars in the bulge 
(here $x_{IMF}$ is set equal to 1.1 for the bulge whereas 
for the solar neighbourhood we adopt a Scalo IMF).}
\end{figure}

\subsection{The [el/Fe] versus [Fe/H] relations}
Here we have computed the temporal evolution of the abundances of 
twenty elements (H, $^{2}$H, $^{3}$He, $^{4}$He, $^{7}$Li, $^{12}$C, $^{13}$C, $^{14}$N, $^{16}$O, {neutron rich}, $^{20}$Ne, $^{24}$Mg, $^{28}$Si, $^{32}$S, $^{40}$Ca, $^{56}$Fe, Ni, Cu, Zn, Kr) but we will not 
discuss them all in detail. 
Predictions on abundance ratios which should be expected in bulge 
stars are shown in Figg.2-3-4-5 (where a comparison is made with the 
same theoretical ratios in halo and disk stars 
belonging to the solar vicinity). 
All ratios are normalized to the solar ones given by 
Anders \& Grevesse (1989). Note that the Mg yields from massive stars adopted here (Woosley \& Weaver, 1995) are probably too low: in fact, the theoretical curve [Mg/Fe] does not fit the observational points in the solar 
neighbourhood if this ratio 
is normalized using the solar abundances measured by 
Anders \& Grevesse and not those predicted by the model itself 
(see also Chiappini et al.\,, 1997 and references therein).

\begin{figure}
\centering
\vspace{12.0cm}
\caption[]{\label{fig:fig 3} The predicted [el/Fe] vs [Fe/H] relations for
$^{16}$O, $^{20}$Ne and $^{24}$Mg in the Galactic bulge (continuous lines) and 
in the solar neighbourhood (dashed lines).}
\end{figure}

\begin{figure}
\centering
\vspace{12.0cm}
\caption[]{\label{fig:fig 4} The predicted [el/Fe] vs [Fe/H] relations for
$^{28}$Si, $^{32}$S and $^{40}$Ca in the Galactic bulge (continuous lines) and 
in the solar neighbourhood (dashed lines).}
\end{figure}

\begin{figure}
\centering
\vspace{12.0cm}
\caption[]{\label{fig:fig 5} The predicted [el/Fe] vs [Fe/H] relations for
Cu, Zn, Ni in the Galactic bulge (continuous lines) and 
in the solar neighbourhood (dashed lines).}
\end{figure}

At present, observational constraints for the bulge stars
are available only for few elements in the range 
-\,1 $\leq$ [Fe/H] $\leq$ +\,0.45 dex (McWilliam \& Rich, 1994). 
In particular, the observed [Mg/Fe] is overabundant by +\,0.3 $\pm$ 0.17 dex relative to solar over almost the full [Fe/H] range; in contrast, Ca and Si closely follow the normal trends for disk giants. We would emphasize 
that there is a qualitative agreement between these observations and theoretical trends as predicted by our model.\\
Note that good models for the bulge in 
Matteucci \& Brocato (1990) give [$\alpha$/Fe] ratios of 
the order of +\,0.5 -- +\,0.4 dex ($\alpha$ indicates O, Mg, Si); 
here we distinguish $\alpha$-elements in two groups showing different trends:
({a}) O, Ne, Mg and ({b}) Si, S, Ca. 
The elements in group (a) show a larger overabundance relative to Fe relative
to those of group (b) and their ratios relative to iron continuosly decrease over the
[Fe/H] range whereas the ratios relative to iron of the elements of group (b) show more constant values.
These differences could be explained in terms of different time scales 
of element production: O, Ne, Mg are produced mostly in 
massive stars on timescales of the order of several million years
whereas Si, S, Ca are produced by both massive stars and binary systems ending their life as SNeIa on timescales varying from 30 million years to several billion years thus behaving more similarly to Fe which originates 
primarily in SNIa events but is also produced by type II SNe. 
Therefore, O, Ne, Mg are overabundant with respect to Fe until SNeIa 
begin to restore the bulk of iron whereas the ratios of
Si, S, Ca relative to Fe have a more flat trend over all the metallicity range
because in explosive nucleosynthesis 
from type Ia SNe there is a non-negligible production of Si, S, Ca 
(Nomoto et al.\,, 1984).

\begin{figure}
\centering
\vspace{12.0cm}
\caption[]{\label{fig:fig 6}
The predicted type Ia SN rates (century$\,^{-1}$) in the Galactic 
bulge (continuous line) and in the solar vicinity (dashed line). 
It is worth emphasizing that the peak occurring in the bulge at 
$t \sim 0.4$ Gyr is mostly due to the shorter time scale of collapse 
and more efficient SFR.} 
\end{figure}

\begin{figure}
\centering
\vspace{12.0cm}
\caption[]{\label{fig:fig 7}
The shorter evolution of the bulge (R = 2 Kpc) with respect to that 
of the solar neighbourhood (R = 10 Kpc) results in an accelerated 
consumption of deuterium, whose abundance quickly falls to zero. 
Moreover, the model predicts minor deuterium depletion in the outer 
regions of our galaxy (R = 16 Kpc is the case sketched here), 
due to longer time scale of collapse and 
consequently low efficiency of star formation.}
\end{figure}

Matteucci \& Brocato (1990) have already stressed that different evolutionary histories lead to different age-metal-licity relations. We confirm their prediction: the point at which the slope of [O, Ne, Mg/Fe] vs [Fe/H] starts changing drastically is attained, in the solar vicinity, at [Fe/H] $\sim$ -\,1.0, corresponding at $t \sim 1$ Gyr. In the bulge, due to faster evolution and larger number of massive stars belonging to 
a single stellar generation, the same point is achieved 
when [Fe/H] $\sim$ +\,0.1, which corresponds 
roughly to $\sim$ 0.4 Gyr. In fact, the central region is 
likely to have undergone a formation 
process so fast 
that type Ia SNe did not have time to  restore the 
bulk of iron until a 
metallicity of $\sim$ +\,0.1 was reached. In addition, the larger number of
massive stars in the bulge relative to the solar neighbourhood leads to larger
[$\alpha$/Fe] ratios, as it is evident from figure 3. 

In Fig.6 we show the rate at which type Ia SNe explode in the Galactic 
bulge compared with that in the solar neighbourhood as 
predicted by our models. 
We can immediately see that in the bulge a peak appears at 400 million years, 
while in the solar vicinity the type Ia SNe rate becomes different 
from zero after $\sim$ 1 Gyr and then increases only slightly 
until the present time.\\
So the time at which type Ia SNe restore the bulk of iron is 
greatly dependent on SFR and IMF 
parameterization: to assume this time scale to be equal to 1 Gyr 
is a proper choice only if we are referring to 
the solar neighbourhood evolution. Otherwise, a different value has 
to be used, according to the assumed 
SFR and IMF prescriptions (Matteucci, 1996). 

The abundance ratios versus [Fe/H] for the other heavy isotopic 
species treated here 
are also shown in Figg.2-5. The differences produced with respect 
to the trends predicted for the solar 
cylinder could 
likely be explained just in terms of different evolutionary histories and greater IMF richness in 
massive stars in the bulge. We do not discuss the trends for Cu, Zn and Ni
because they have been studied already in Matteucci et al. (1993). We only remind here that all these elements are assumed to be mostly produced by SNeIa,
although Cu and Zn are partly produced as s-process elements.

\subsection{Light elements in the bulge}
\subsubsection{Deuterium}
In the scenario of a shorter accretion process and stronger star formation 
rate assumed here for the Galactic bulge it is interesting to have a 
look at the temporal trend exhibited by the deuterium abundance 
and compare it with the same quantity as predicted 
for the solar neighbourhood 
under appropriate assumptions on SFR and IMF.\\
Deuterium is only destroyed in stellar interiors and  its abundance 
in the central regions 
of the Galaxy is expected to
fall quicker  due to the faster stellar consumption with respect 
to the solar neighbourhood. In Fig.7a the deuterium evolution 
at three different galactocentric distances is shown; 
the deuterium abundance  at the distance representative of the bulge 
falls by almost three orders of magnitude in a time scale as 
short as $\sim$ 4 Gyr. 
Here  
we adopt a primordial deuterium abundance by mass of 6 $\times$ 10$\,^{-5}$ 
corresponding to the deuterium value of  
Burles and Tytler (1997) although
the primordial deuterium abundance, 
as deduced from  chemically unevolved high redshift clouds, 
is rather controversial, differing by one order of magnitude
according to different sources. 
The revised abundance of Burles and Tytler (1997) is  
(D/H) = 3.3 $\times$ 10$\,^{-5}$ but  in the system studied by Webb et al.\,
(1997) it is of 2 $\times$ 10$\,^{-4}$. Recently, Tosi et al. (1998)
showed that realistic models of Galactic chemical evolution favor the value
of Burles and Tytler  (1997).
The deuterium astration 
versus metallicity is shown in Fig.7b. It is remarkable that there is no significant difference among the curves at
2, 10 and 16 Kpc implying a very similar behaviour for the deuterium evolution between the bulge and the solar cylinder. What differs is the time scale
of the whole process which is very short in the bulge 
as compared to that of the Galactic disk for producing the same amount of 
astration. The present time deuterium abundance in the bulge should 
be about three orders of magnitudes smaller 
than that in
the disk. 
However, this prediction would  hardly be verified since direct 
deuterium observations in the bulge cannot avoid 
the local disk contamination along the line of sight.
The interesting aspect of 
the analysis presented here for the bulge is that it shows that the
large difference in  the deuterium abundances measured in high redshift clouds 
can  hardly  be ascribed to different phases of bulge-like evolution for 
these clouds. In fact, the deuterium astration  is firmly bounded by
the increase in the metallicity. The high redshift clouds used for
the deuterium determination have
always a rather low metallicity around [Fe/H] $<$ -\,2 and therefore, 
according to the deuterium chemical evolution discussed here, they should have suffered a similar amount of astration.   

\subsubsection{Lithium}
We have calculated in detail the
temporal evolution of lithium ($^{7}$Li) in the bulge. 
Minniti et al.\,(1998) have recently measured for the first time 
the $^{7}$Li abundance in the atmosphere of a bulge 
star,
which is the source star of the MACHO microlensing event 97\,BLG\,45.
The star is a dwarf and Minniti et al. have obtained log N$_{Li}$\footnote{log N$_{Li}$ = log$_{10}$ (X$_{Li}$/7\,X$_{H}$) + 12, where X$_{el}$ is the mass fraction in form of the generic element $el$.} = 2.25 $\pm$ 0.25. The authors suggested that this value is consistent 
with the most recent determination of the  primordial lithium abundance from 
the halo dwarf population (Bonifacio \& Molaro, 1997). However we argue below
that this is probably a coincidence and  that the Li abundance in this star
could not be indicative of the Li primordial value.  

We want first to remind  to the reader  
the essential of the lithium evolution in the solar neighbourhood 
as discussed in  Matteucci et al.\,(1995). 
These authors presented Galactic chemical evolution 
models which took into account 
recent developments about 
$^{7}$Li stellar production and concluded  that the rise off from
the plateau in the diagram log N$_{Li}$ -- [Fe/H] is best  
reproduced with a mix of $^{7}$Li production by $\nu$-supernovae, 
carbon stars and massive asymptotic giant branch (AGB) stars.

We have computed here four models (see Table 1)
for the lithium abundance evolution of the bulge, changing 
the prescriptions about $^{7}$Li stellar production. 
In all the models here we assume an 
AGB production of lithium dependent on the initial 
metallicity of the progenitor. This effect is 
parameterized through the value of M$_{up}$ - the 
maximum mass of a star which develops a degenerate 
CO core after the exhaustion of He in its center - 
following 
Matteucci et al.\,(1995), who in turn based their choice 
on the calculations of Tornamb\'e \& Chieffi (1986). 
The overall effect of such a parameterization is 
that low metallicity AGB stars do not contribute 
any lithium.\\
We want to recall that lithium enrichment by AGB stars occurs both through stellar winds 
and planetary nebula ejection (see Matteucci et al.\,, 1995, and references therein). 
The $^{7}$Li production in type II SNe occurs by 
means of the $^{4}$He\,($\nu$\,,\,$\nu$ n)\,$^{3}$He\,($\alpha$\,,\,$\gamma$)\,$^{7}$Be\,(e$^{-}$\,,\,$\nu_{e}$)\,$^{7}$Li reaction, as discussed in 
Woosley et al.\,(1990). Note that the yields we have adopted here 
for these objects 
(Woosley \& Weaver, 1995) are metallicity dependent.

The models inputs are listed in Tab.1 and the data sets relative to
the solar neighbourhood are from 
Boesgaard \& Tripicco (1986), Lambert et al.\,(1991), Pasquini et al.\,(1994) 
and from the compilation of Delyiannis et al.\,(1990)
for stars with [Fe/H] $>$ -\,1.5 and T$_{eff}$ $>$ 5500 K 
and from Bonifacio \& Molaro (1997) for the halo stars. 
The primordial Li value is assumed 2.2 close to the recent determination 
of Bonifacio \& Molaro (1997).\\

\begin{figure}
\centering
\vspace{12.0cm}
\caption[]{\label{fig:fig 8}
Log N$_{Li}$ -- [Fe/H] diagram showing the observations of several authors relative to the solar neighbourhood. The curve shows the best fitting chemical evolution model (prescriptions on stellar lithium production like in model 1).}
\end{figure}

In figure 8 we show the predicted log N($^{7}$Li) versus [Fe/H]
when the 
Matteucci et al.\,(1995)  Li best model prescriptions are included in the Chiappini et al. model for the solar vicinity. 
The figure shows that the agreement with the observations is quite good since we want to reproduce the upper envelope of the data distribution.
The main difference between the chemical evolution model adopted by Matteucci et al.
(1995) and Chiappini et al. (1997) adopted here is in the fact that the latter considers the evolution of halo-thick disk separated from that of the thin disk,
but this clearly does not affect the predicted behaviour of Li versus Fe.

\begin{figure}
\centering
\vspace{12.0cm}
\caption[]{\label{fig:fig 9}
Log N$_{Li}$ -- [Fe/H] theoretical relations in the Galactic bulge (continuous lines) and in the solar neighbourhood (dotted lines) as predicted by our models in four cases: a) only astration is acting in stellar interiors; b), c) and d) stellar production by different categories of objects is considered (prescriptions on stellar Li synthesis like in model 1, Tab.1).}
\end{figure}

In Fig.9 are highlighted the contributions of the different 
galactic sources to Li enrichment. Fig.9a shows the Li astration in absence of Li production,
while Fig.9b and Fig.9c show the contribution for the AGB stars and 
$\nu$-supernovae. The difference between the 
two AGB curves reflects the faster 
iron enrichment in the bulge in relation to the time scale of the AGB Li 
production. If AGB stars would be the only Li sources then the Li evolutionary
 curves for the bulge and for the disk would be different with the plateau 
extending towards higher metallicities in the bulge. 
In Fig.9d all the sources are considered giving a prediction for the present
Li abundance in the bulge of $\simeq$ 3.5, 
which is considerably higher than that of the disk.

Minniti et al. did not provide the iron abundance for their star. 
However by using their equivalent width of FeI 6705.1 
line we estimated a metallicity  of  around [Fe/H] $\approx$ +\,0.4 dex. This 
abundance is obtained with a model of T$_{eff}$ = 6000 K, i.e. the temperature 
adopted by Minniti et al.\,, log g = 4,  microturbulence of 1.7 km s$\,^{-1}$ 
and solar opacity distribution functions (ODF).   
With this entry the Li evolutionary curve  which well reproduces the 
observed trend in the solar neighbourhood fails 
to reproduce the Minniti star in the bulge (curve 1 in Fig.10).

By changing the $^{7}$Li production in stars is possible  to 
extend the plateau further towards higher metallicities, but 
hardly  long 
enough to approach  the experimental point of Minniti et al.\\
For instance, a possible solution may be to 
reduce the $^{7}$Li contribution by AGB stars;   
alternatively one can  suppress a particular category of $^{7}$Li factories. 
Curves 2 and 3 result from the  hypothesis that $\nu$-supernovae do not contribute any lithium 
to the interstellar medium pollution; 
model 3 adopts also a lower 
$^{7}$Li abundance in the atmosphere of the Li-rich AGB stars. 
A further
improvement can be achieved by lowering  the IMF slope. Curve 4 shows the results of model 4
with a slope for the IMF of 1.1.

Therefore to consider the lithium abundance measured in the atmosphere of 
97\,BLG\,45  
as representative of the primordial value requires a substantial  revision 
of the Li evolutionary behaviour which, on the other hand, is found 
to match satisfactorily  the Li observed trend in the solar vicinity.
Otherwise - and this is our claim - 
the $^{7}$Li abundance of 97\,BLG\,45 has suffered some  $^{7}$Li depletion by one of those Li depletion mechanisms which are known to act  during 
the pre or main sequence in stars 
of relatively high metallicities.
Several different mechanisms have been proposed such as 
diffusion (Michaud, 1986),
turbulent mixing (Charbonnel \& Vauclair, 1992), 
mixing driven by rotation (Pinsonneault et al.\,, 1998), 
pre-main sequence depletion (Ventura et al.\,, 1998).
Metal rich stars have deeper convection zones compared to the metal weak ones
making relatively easier the mixing  of the atmosphere with the
more internal and   hot layers where Li is burned  (T $>$ 2 $\times$ 
10$\,^6$ K). 
Considering the field stars with solar or moderate metal deficiency 
shown in Fig.8 one sees that a considerable fraction of these stars 
shows a Li abundance
close to the primordial value. 
For some of these stars with available Be abundance observations 
it is possible to
have indirect evidence that Li has been depleted (Molaro et al.\,, 1997) 
implying that their original Li content was much higher than the 
primordial value.
Since in the bulge the 
metal rich stars  are older than the  solar neighbourhood ones 
with similar metallicity,  the probability of Li depletion should be
higher in the bulge stars than in the disk stars.

The predicted Li evolution curve with respect to the metallicity
for the bulge is mimicking that for the solar neighbourhood 
(Fig.9d). However it will be very difficult to test this observationally. 
In fact,  observations
in the bulge will select preferentially metal rich stars and it will be
particularly difficult to pick-up a metal poor star belonging to the bulge
as it is for instance for the galactic thick disk. Only observations of 
metal poor dwarfs in the bulge will be informative of the initial Li content 
before the significant and quick Li production in the bulge.

\begin{figure}
\centering
\vspace{12.0cm}
\caption[]{\label{fig:fig 10} 
Log N$_{Li}$ vs [Fe/H] relation as predicted by our models 
for the Galactic bulge. The diamond is the bulge star 97\,BLG\,45.} 
\label{fig:fig 10}
\end{figure}

\section{Conclusions}
In this paper we have computed chemical evolution models for the Galactic bulge. We have shown that a scenario 
characterized by an evolutionary process much faster than that in the solar neighbourhood and even faster than that in the halo 
(see also Renzini, 1993) allows us to reproduce very 
well the observed metallicity distribution of bulge K giants.\\
We have stressed that to fit the most recent observed distribution (McWilliam \& Rich, 1994) 
one needs to choose an IMF with a power index in the range 1.1 -- 1.35, that is to say that 
in the bulge the formation of massive stars relatively to the solar vicinity region should be favored.\\
Notwithstanding the uncertainties still present in the remaining 
model parameterization (SFR efficiency, 
time scale of collapse, nucleosynthesis prescriptions) we think that 
our previous considerations 
should be considered as a probe of the necessity of a variation in 
the IMF slope between 
the bulge and the solar neighbourhood.\\
This requirement stands out also in the work of Matteucci \& Brocato (1990). 
They used values of $\nu$ 
and $\tau$ slightly different from those adopted here 
($\nu$ = 10 Gyr$\,^{-1}$, $\tau$ = 0.01 Gyr) 
but reached the same conclusions: the position of the metallicity 
peak can be reproduced only by allowing 
that more massive stars have been formed in the bulge than in 
the solar vicinity. Our results also show that there is no need
for a pre-enriched gas from the halo to form the bulge. \\
We have then explored the effect of the faster bulge formation
and flatter IMF 
on the temporal evolution 
of several chemical and isotopic species. We have made theoretical 
predictions which need to be confirmed or 
disproved by future observations. 
In particular, we predicted that the $\alpha$-elements should 
show an overabundance 
relative to iron for most of the [Fe/H] range, at variance 
with what happens in the solar neighbourhood.
This overabundance is not constant for all the $\alpha$-elements 
but it depends, 
for a specific element, on the relative production of this element 
by supernovae of different types.
In particular, [Si/Fe], [S/Fe] and [Ca/Fe] should be less overabundant
and more constant than [O/Fe], [Mg/Fe] and [Ne/Fe].
The [$^{12}$C/Fe] ratio is predicted to be solar for the whole 
range of metallicity as in the solar vicinity,
whereas [$^{13}$C/Fe] and [N/Fe] are predicted to be higher than in the solar neighbourhood at low metallicities 
as due to the faster star formation rate  adopted for the bulge. On the other hand, due to the larger number
of massive stars relative to the low and intermediate mass stars assumed for the bulge, these two ratios 
are lower at solar metallicities.

We also predicted that deuterium should be completely exhausted in the bulge, 
due to the intense star formation, 
and the $^{7}$Li abundance versus [Fe/H] trend should be similar to what is found in the solar vicinity (Fig.9d).
In order to reproduce the low Li abundance recently 
claimed by Minniti et al.\,(1998)
for a main sequence bulge star, under the assumption that it reflects the original Li abundance in the gas,
one has to suppress one or more sources of stellar lithium. 
On the other hand, it is likely that a star of such a high metallicity ([Fe/H] $\sim$ + 0.4 dex) would have depleted most of its Li content already during the pre-main sequence phase (Ventura et al.\,, 1998) although
more measurements of Li abundance in bulge stars are necessary 
before drawing firm conclusions.

\begin{acknowledgements}
We would like to thank C.\,Chiappini for useful suggestions
and P.\,Ventura for calculating a stellar evolution model.
\end{acknowledgements}

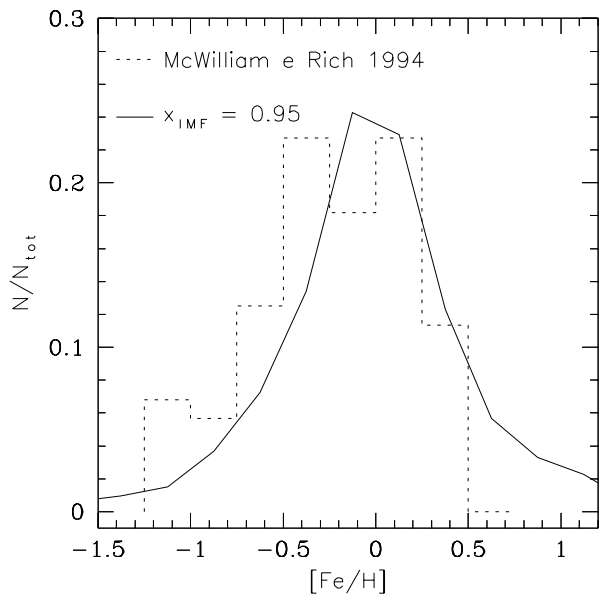
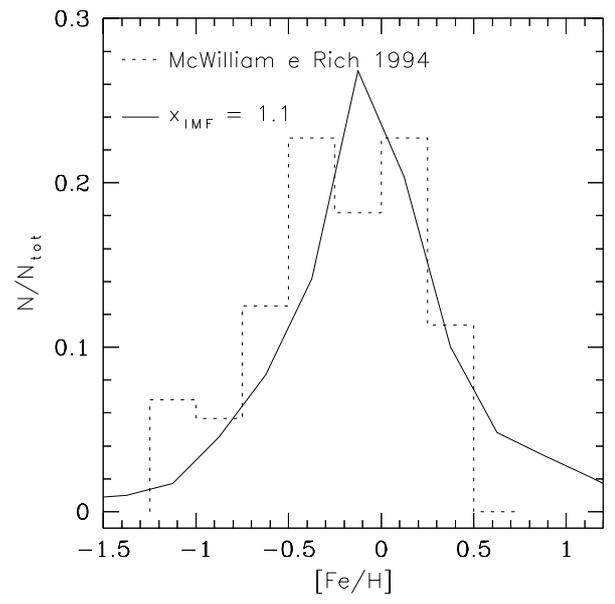
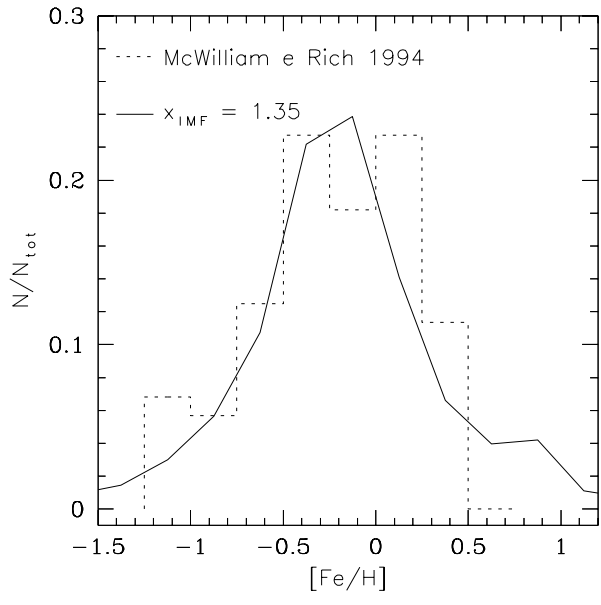
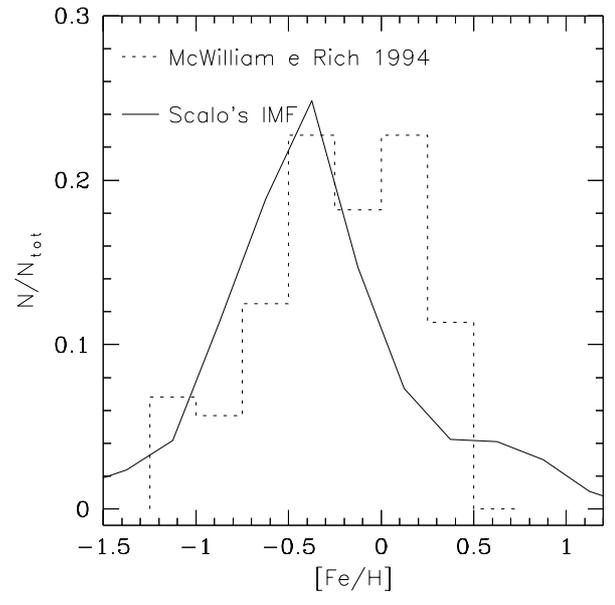
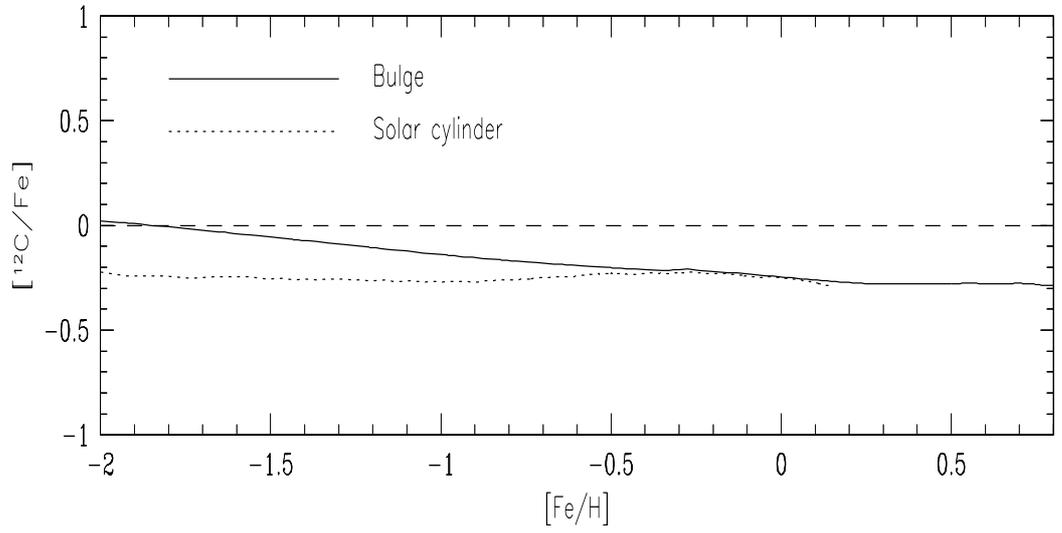
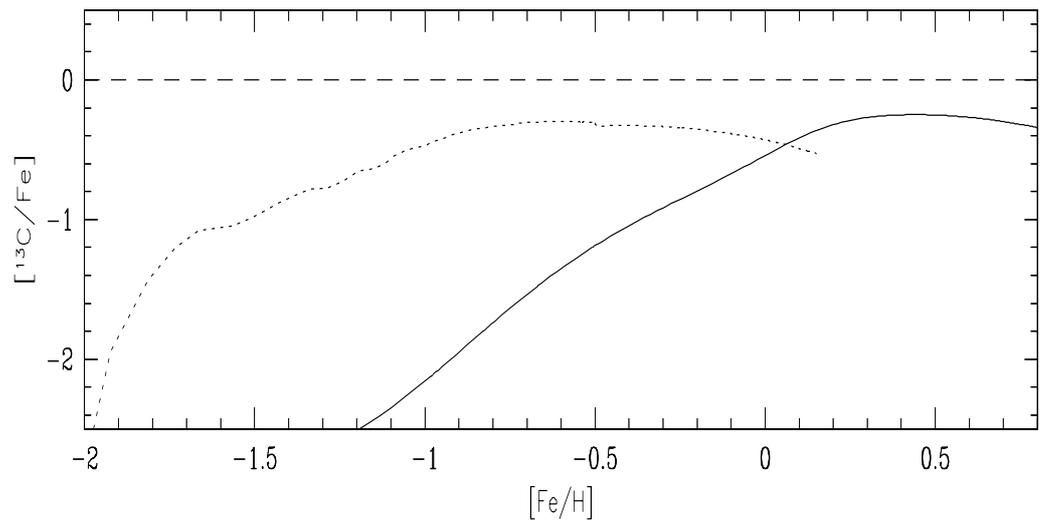
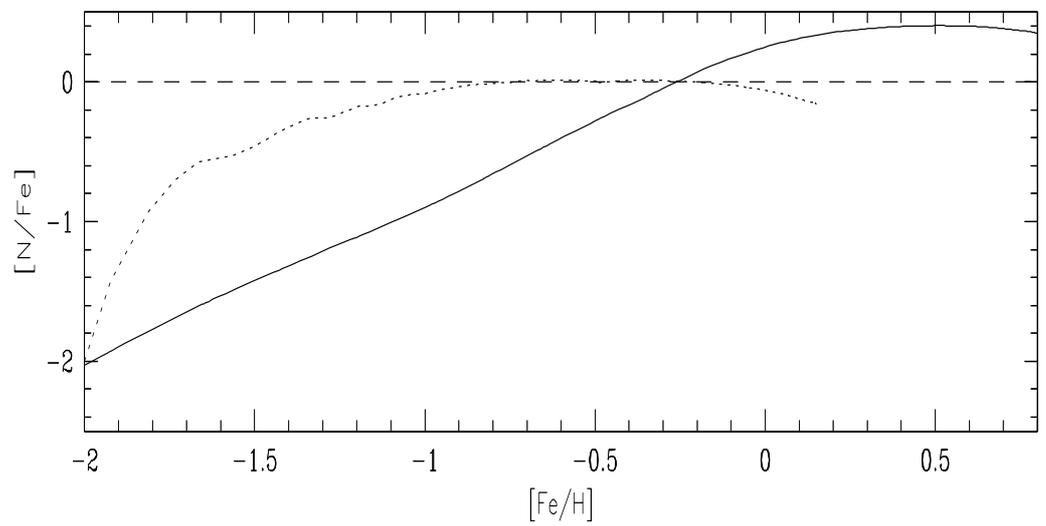
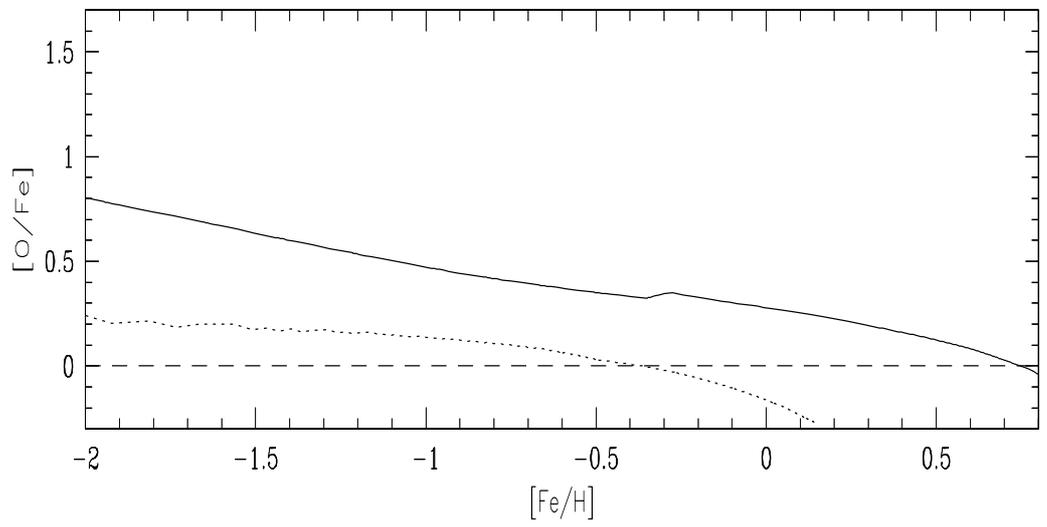
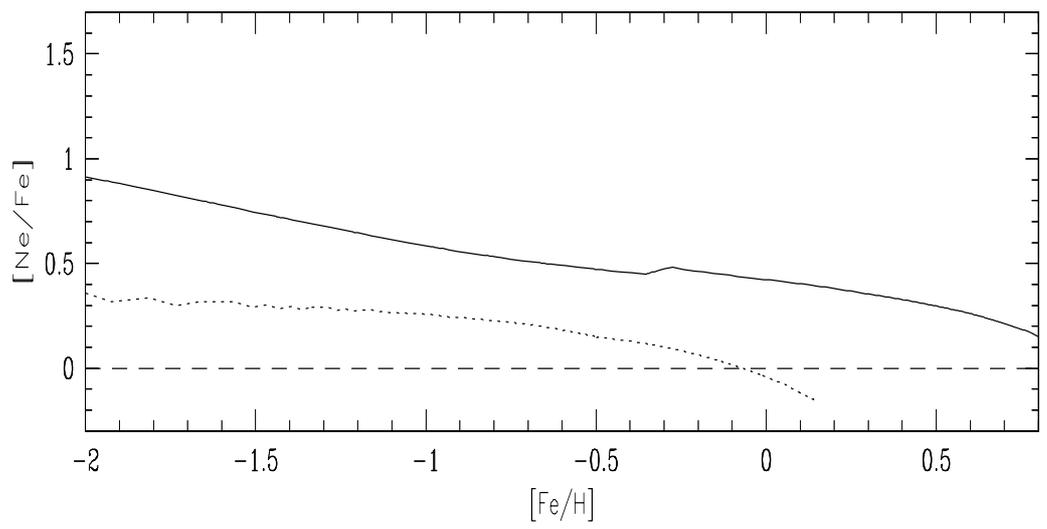
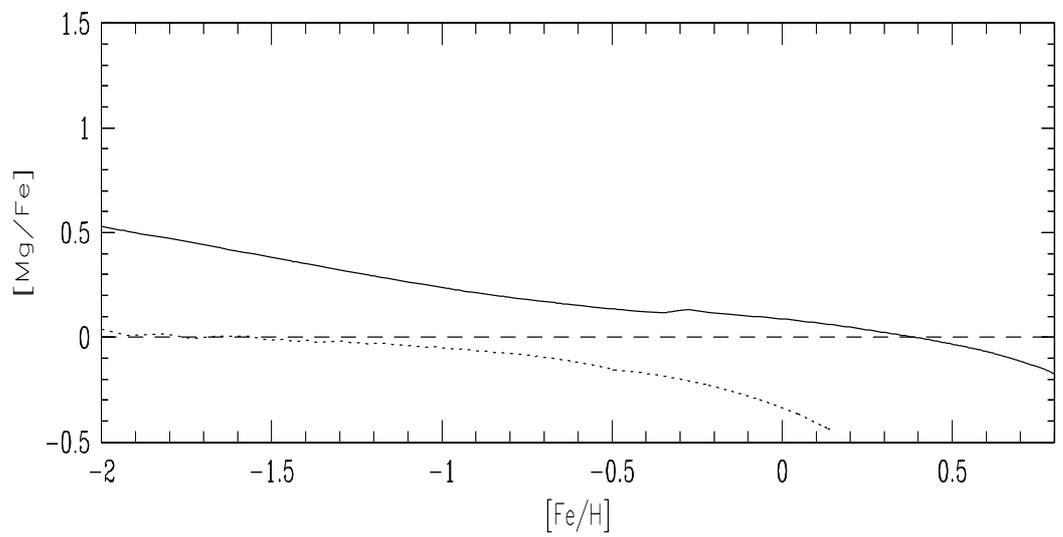
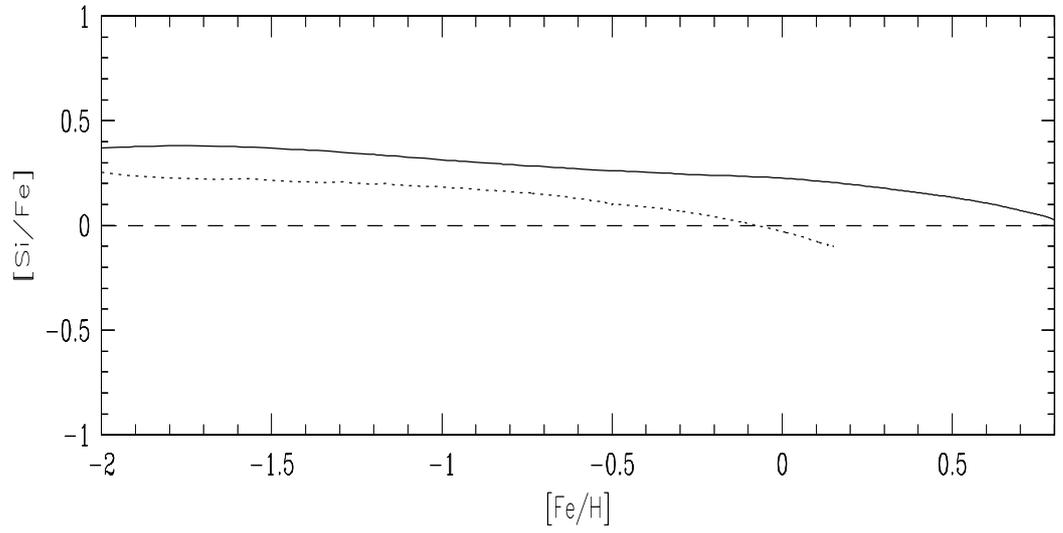
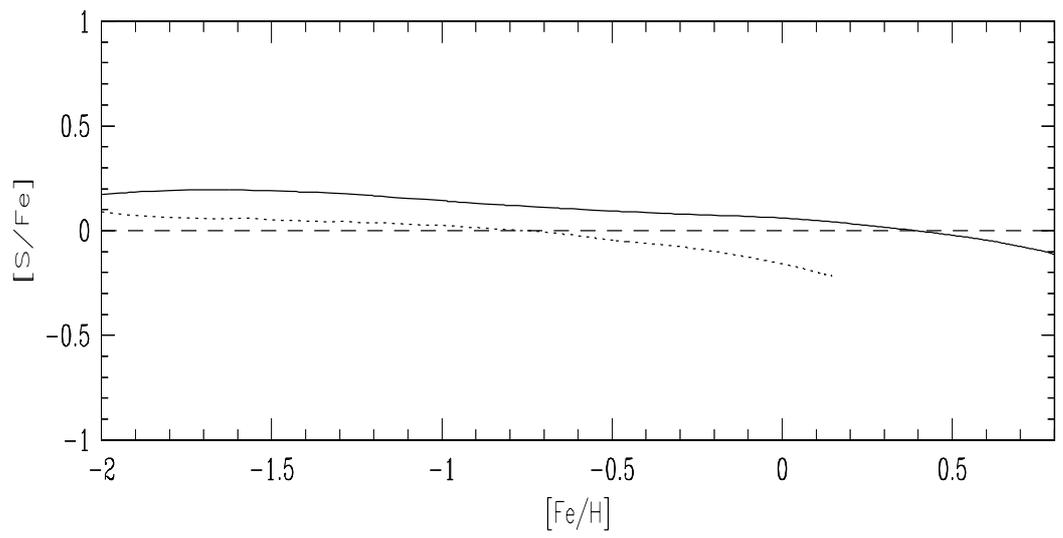
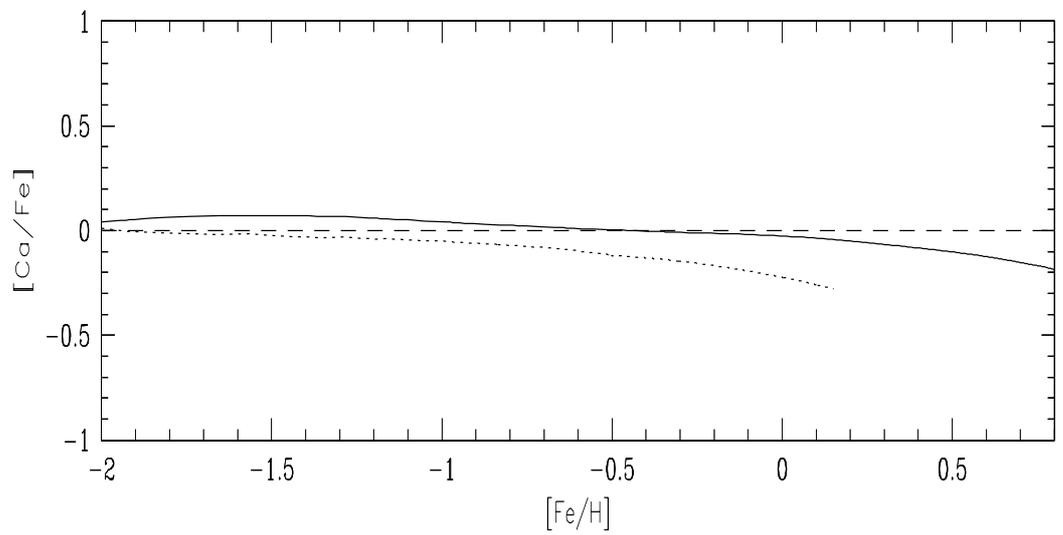
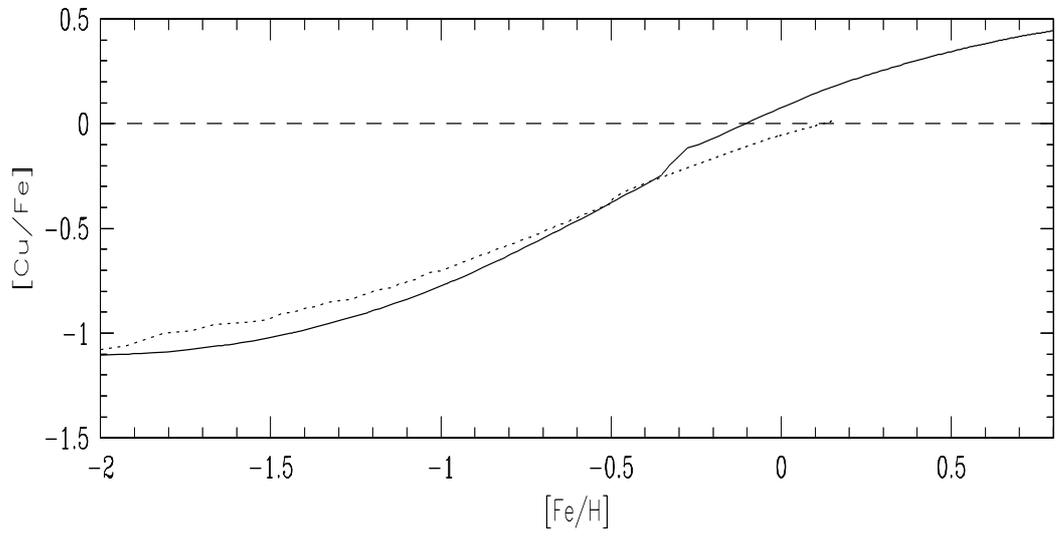
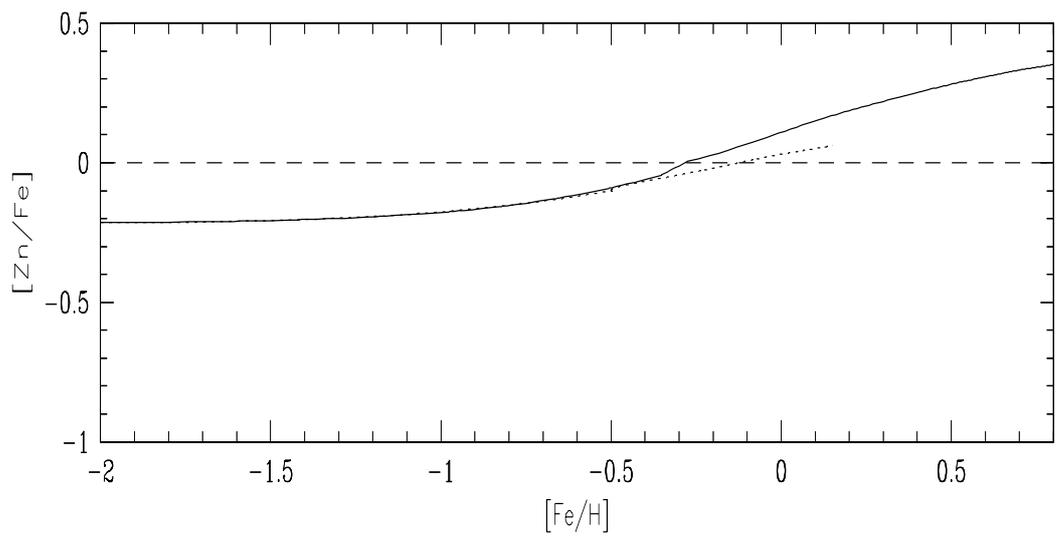
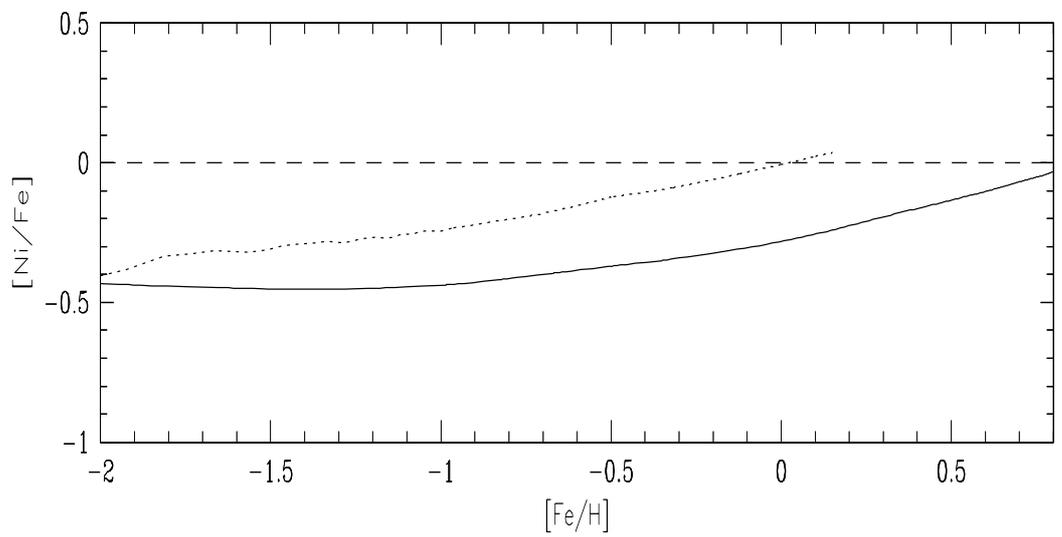
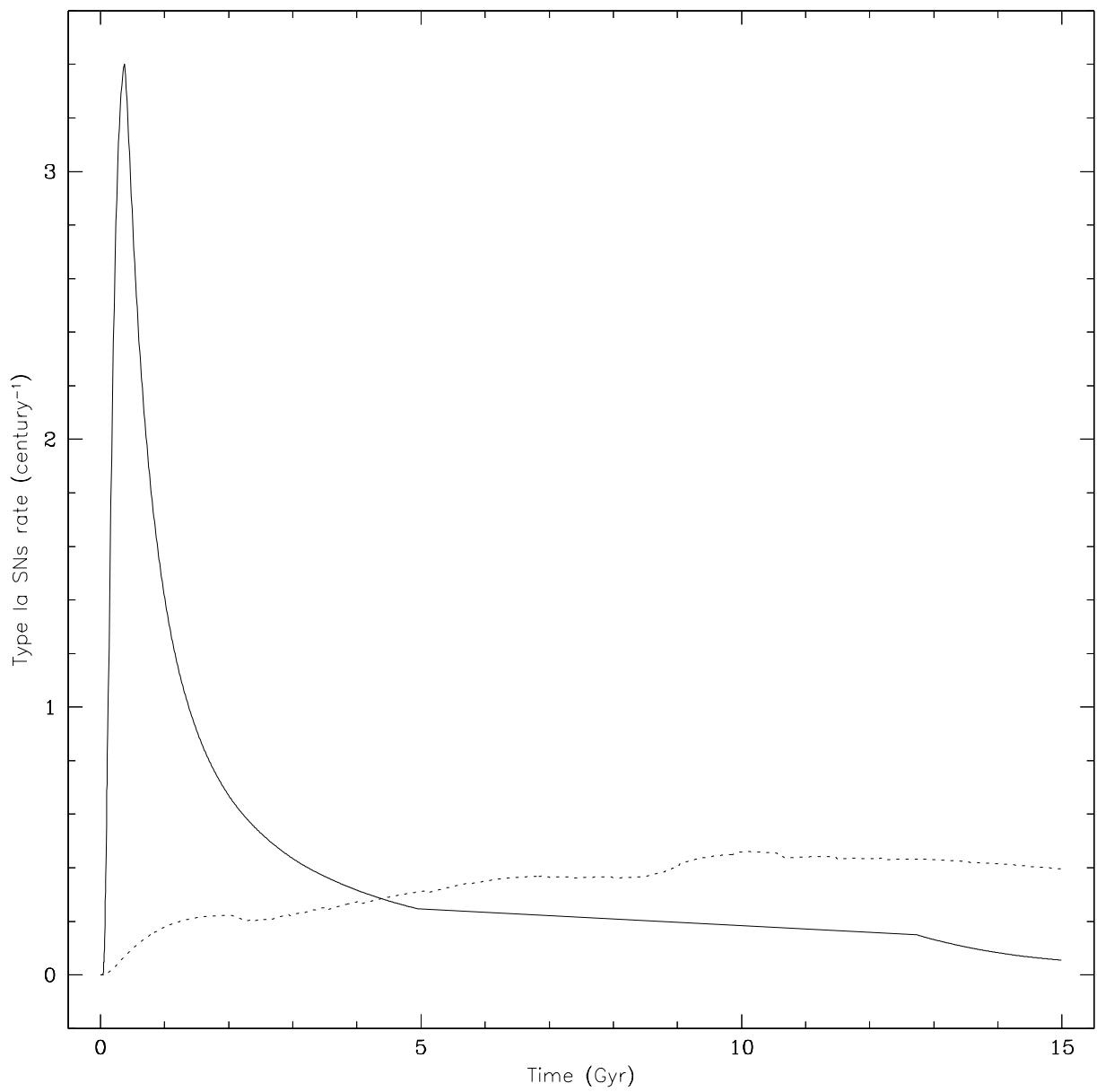
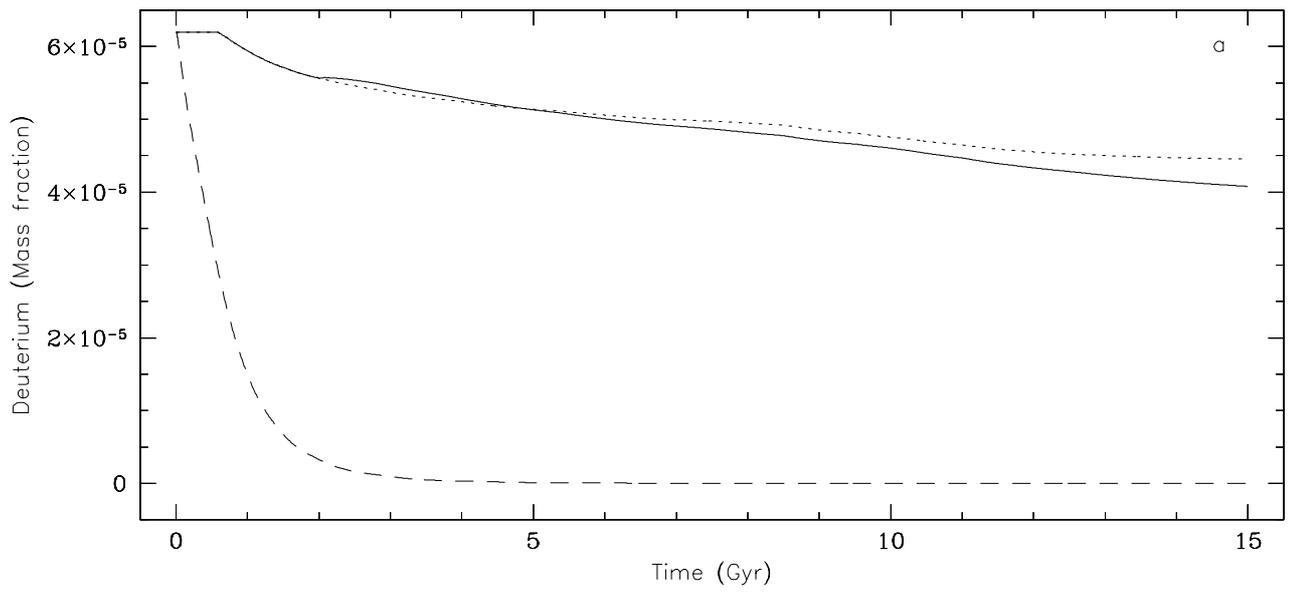
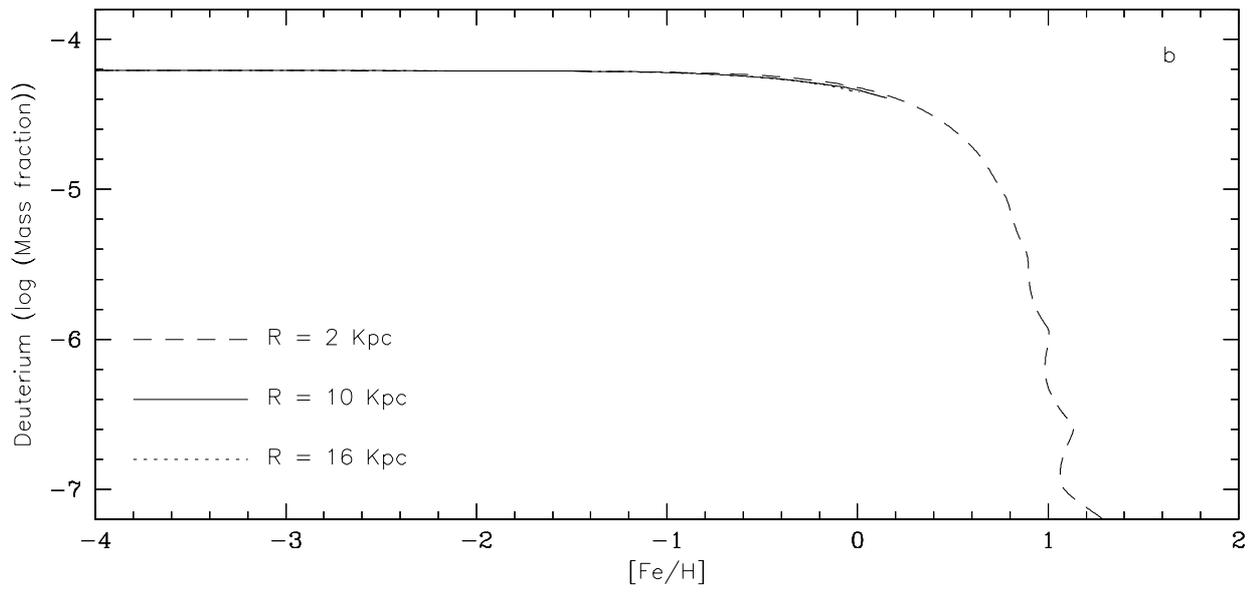
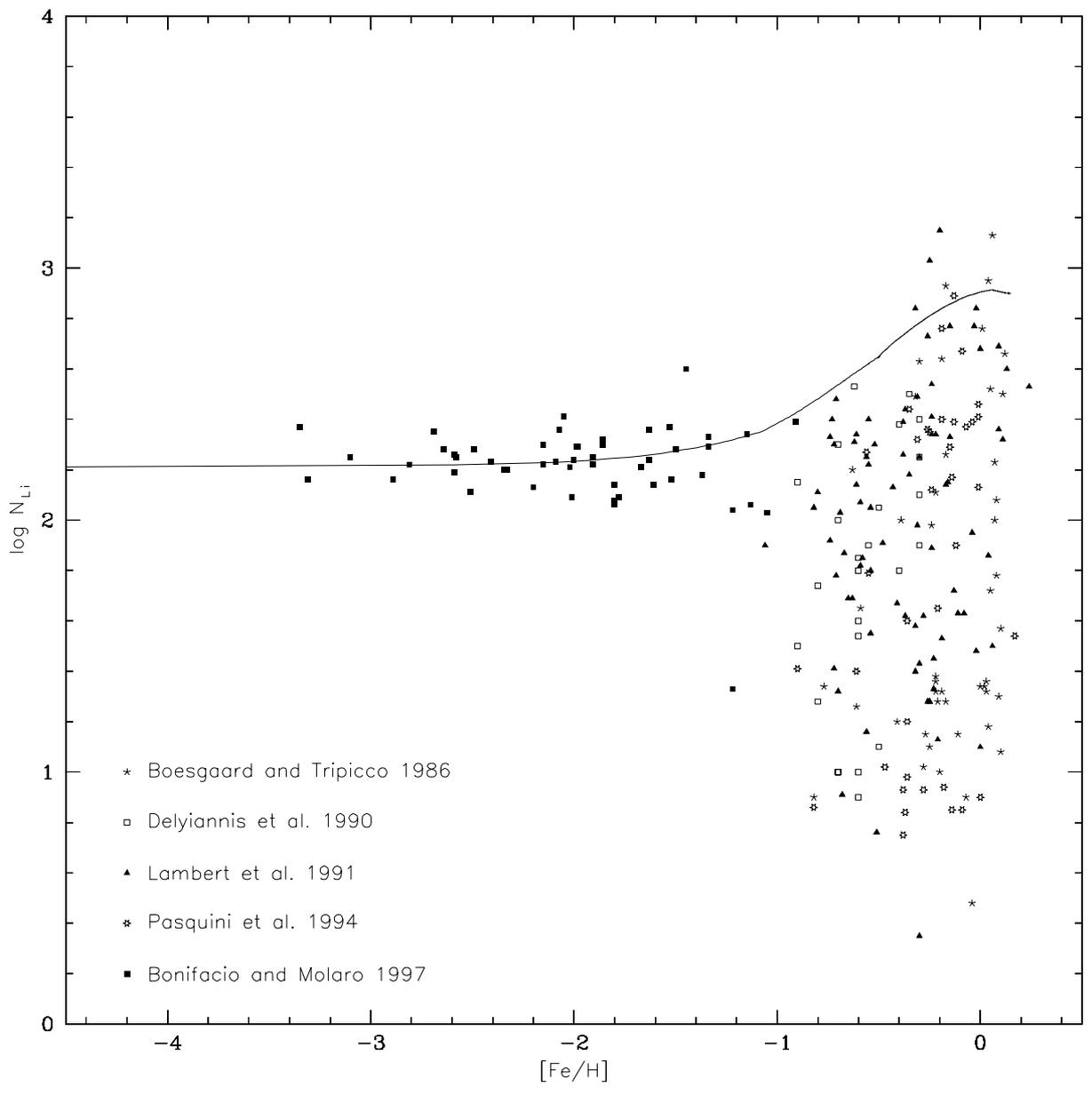
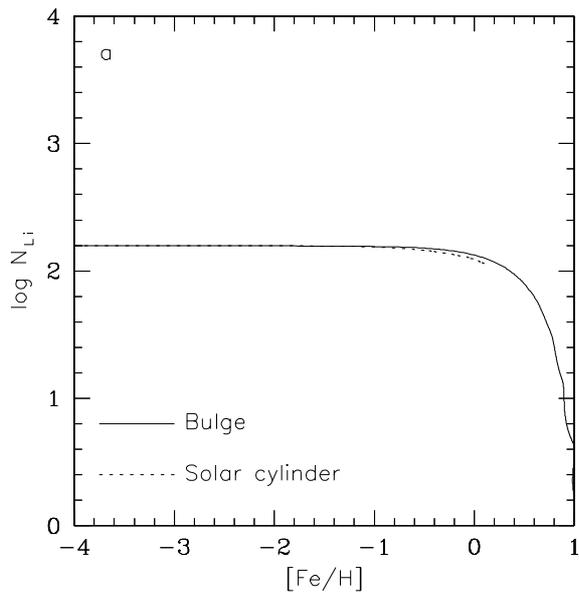
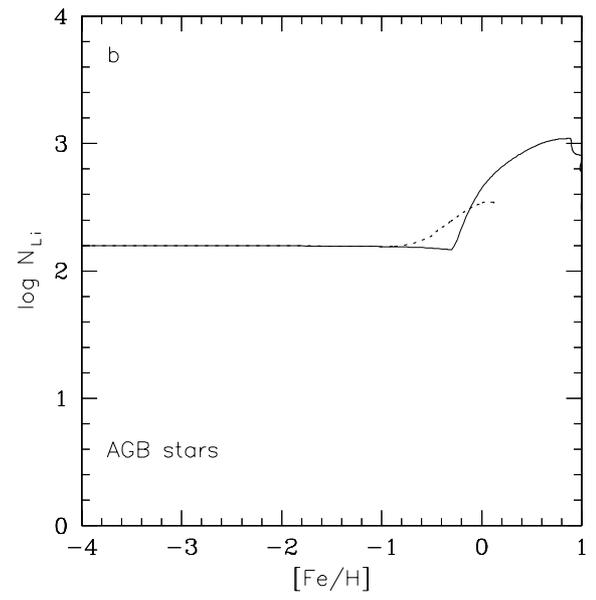
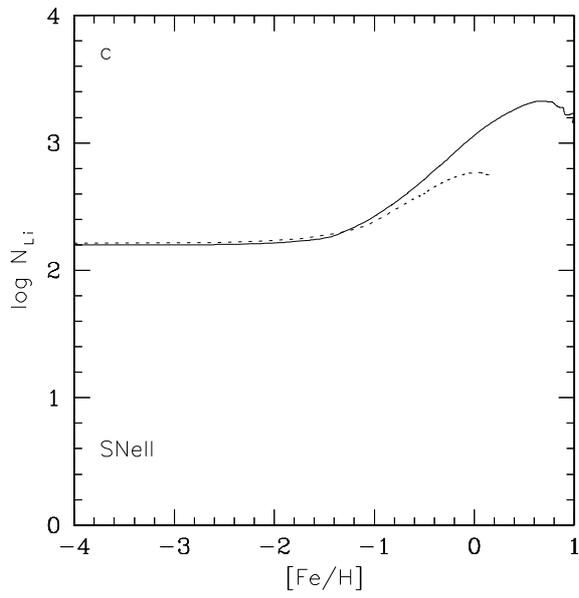
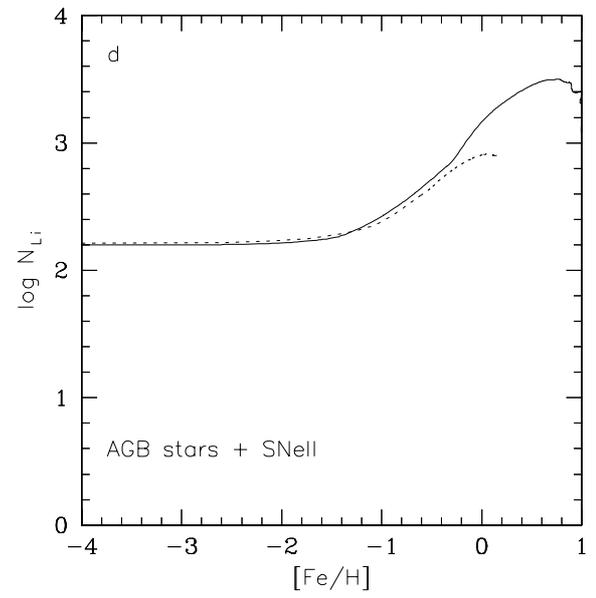
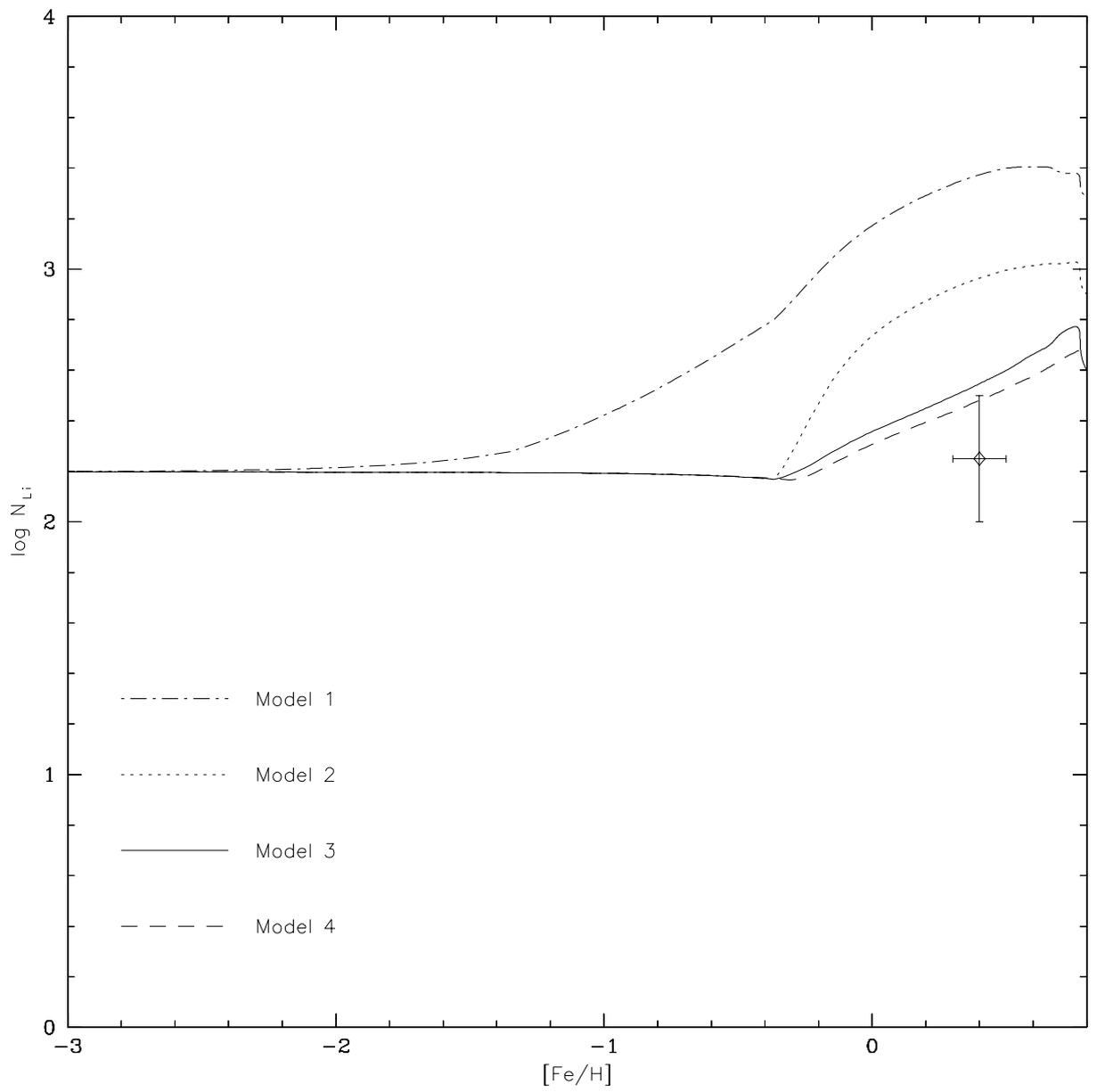
\end{document}